\documentclass[aps,showpacs,twocolumn,showkeys,amsmath,amssymb,floatfix,nofootinbib,preprintnumbers]
{revtex4}
\usepackage{bm}
\usepackage{epsfig}

\begin{document}

\title{A New Picture for the Formation and Decay of the Exotic $XY \! Z$
Mesons}

\author{Stanley J. Brodsky}
\email{sjbth@slac.stanford.edu}
\affiliation{SLAC National Accelerator Laboratory, Stanford
University, Stanford, California 94039, USA}

\author{Dae Sung Hwang}
\email{dshwang@sejong.ac.kr}
\affiliation{Department of Physics, Sejong University, Seoul 143-747,
South Korea}

\author{Richard F. Lebed}
\email{richard.lebed@asu.edu}
\affiliation{Department of Physics, Arizona State University, Tempe,
Arizona 85287-1504, USA}

\date{June, 2014}

\begin{abstract}
We present a new dynamical picture that identifies the formation of
the exotic $\bar c c$-containing states $XY \! Z$ with the
confinement-induced hadronization of a rapidly separating pair of
a compact diquark and antidiquark.  This picture combines the
advantages of diquark-based models, which can accommodate much of the
known $XY \! Z$ spectrum, with the experimental fact that such states
are both relatively narrow and are produced promptly.  It also
naturally explains the preference of some of the exotic states to
decay to $\psi(2S)$, rather than $J/\psi$, in terms of a simple
wave-function overlap effect.
\end{abstract}

\preprint{SLAC-PUB-16001}


\maketitle



The 2003 Belle discovery~\cite{Choi:2003ue} of an unusual
charm\-onium-like state, appearing as a resonance of $\pi^+ \pi^-
J/\psi$ and now called $X(3872)$, has led to the observation of
numerous related states during the past
decade~\cite{Brambilla:2014aaa} at Belle, BaBar, CDF, D0, CLEO, CMS,
LHCb, and BESIII.  Indeed, the BESIII experiment continues to present
evidence for new exotic states even in the past few
months~\cite{Yuan:2014rta}.  These states do not fit into the standard 
nonrelativistic quark model of a single $\bar c c$ pair with
separation $r$ interacting via a potential $V(r)$, which had been
successful in accommodating all of the previously known charmonium
states~\cite{Eichten:1978tg,Eichten:1979ms,Barnes:2005pb}; instead
they are believed to be tetraquark ($\bar c c \bar q q^\prime$) states
currently named $X$, $Y$, or $Z$\footnote{In the current nomenclature,
the neutral(charged) states observed in $B$ decays are labeled $X(Z)$,
whereas the $Y$ are the neutral, $J^{PC} = 1^{--}$ states observed in
initial-state radiation $e^+ e^-$ processes.}.  Notable evidence
supporting this identification includes the facts that $X(3872)$ is an
extremely narrow ($\Gamma < 1.2$~MeV) $J^{PC} = 1^{++}$ state but is
tens of MeV lighter than the nearest quark-model candidate
$\chi_{c1}(2P)$~\cite{Barnes:2005pb}, and the recent confirmation at
LHCb~\cite{Aaij:2014jqa} of the charged $J^P = 1^+$ state $Z(4430)$ as
a resonance decaying into $\pi^- \psi (2S)$.  This first verification
of the existence of exotic hadrons, which possess neither meson ($\bar
q q$) nor baryon ($qqq$) valence structure, is an exciting advance for
QCD; a key challenge is to uncover the dynamical structure of these
states.

One can imagine the binding of a ($\bar q_1 q_2 \bar q_3 q_4$) state
to occur in a variety of ways.  First, the four valence quarks can all
interact democratically, which one may call a ``true'' tetraquark.
However, simple SU(3) color group theory shows that the combination of
two quarks (each a color {\bf 3}) and two antiquarks (each a color
$\bar{\bm 3}$) can form an overall color singlet in only two
independent ways---matching the color structure of factorized
two-meson states $(\bar q_1 q_2)(\bar q_3 q_4)$ and $(\bar q_1
q_4)(\bar q_3 q_2)$.  In large $N_c$ QCD, this fact has long been used
to argue that narrow tetraquark states do not occur, since the
four-quark source operators needed to create them are saturated by
two-meson states.  Weinberg has recently
showed~\cite{Weinberg:2013cfa} that this argument contains a loophole;
however, his scenario requires modifications that go beyond the usual
large $N_c$ counting rules and
structures~\cite{Lebed:2013aka,Cohen:2014via,Cohen:2014tga}.

The color factorization property of the tetraquark naively suggests a
two-meson molecule structure for the observed states.  Indeed, many of
the $XY \! Z$ states lie close to such thresholds ({\it e.g.},
$m_{X(3872)} \approx m_D + m_{D^*} \approx m_{J/\psi} + m_\omega$),
suggesting a molecule with a small binding energy $E_b$ via a van der
Waals-type attraction~\cite{Brodsky:1989jd}.  However, the typical
scattering length of such a state is given by its Compton wavelength
$\lambda_C \propto E_b^{-1/2}$, which for some of the observed cases
can be as large as 10~fm in size.  The prompt (pure QCD) experimental
production rate of $X(3872)$ argues against the generation of such
extended states~\cite{Bignamini:2009sk}.  Nevertheless, numerous
papers have argued for the molecular picture (see
Ref.~\cite{Brambilla:2010cs} for many references and extensive
discussion); {\it e.g.}, as a mixture of the $\chi_{c1}(2P)$ and
$(D^{*0} \bar D^0)$.  Alternatively, the $XY \! Z$ states could be
{\it hadrocharmonium}, an ordinary charmonium state embedded in a
light-quark cloud~\cite{Voloshin:2007dx}; however, such states would
presumably mix with conventional charmonium states with the same
quantum numbers.

In this paper, we have been inspired by a well-known hypothesis for
the $XY \! Z$ states---that of a diquark-antidiquark ($\delta$-$\bar
\delta$) pair, $[q_2 q_4][\bar q_1 \bar
q_3]$~\cite{Maiani:2004vq}\footnote{States formed from color {\bf 8}
$[\bar q q]$ diquarks have also been
considered~\cite{Hogaasen:2005jv,Buccella:2006fn}.}.  The principal
advantage of any diquark picture is its flexibility in incorporating
QCD color physics not available in approaches in which the individual
$\bar q q$ pairs are bound into color singlets before any other
effects are considered.  This flexibility of the $\delta$-$\bar
\delta$ picture is also its chief drawback, because one must sort
through many possible dynamical configurations in order to determine
which ones capture the essential physics of the $XY \! Z$ states.
Moreover, the $\delta$-$\bar \delta$ picture supports numerous
hadronic states, due to the spin and flavor degrees of freedom of each
quark; without some simplifying constraint on the allowed dynamics,
the picture rapidly becomes unwieldy and loses predictivity.
Nevertheless, at least one recent
collaboration~\cite{Prelovsek:2014swa} finds that the charged
$Z_c(4020)$ state emerges from lattice simulations only if diquark
interpolating operators are included.

As a simple prototype of a $\delta \bar \delta$ state, consider a QED
{\it tetralepton\/}, formed from $(\mu^+ \mu^- e^+ e^-)$.  Such a
state is not as exotic as it might first appear, since the
dipositronium state $(e^+ e^-)(e^+ e^-)$ has actually been
produced~\cite{Cassidy:2007aa}.  The strong internal Coulomb forces
can be canceled by forming bound, neutral components in two ways,
either as $(\mu^+ \mu^-)(e^+ e^-)$ (a ``true
muonium''\cite{Brodsky:2009gx}-positronium molecule) or $(\mu^+
e^-)(e^+ \mu^-)$ (a muonium-antimuonium molecule); then the strongest
residual force between them is the coupling between their magnetic
dipole moments, which in turn is strongest when each of the neutral
components contains a lighter particle.  Thus, one finds that the
$(\mu^+ e^-)(e^+ \mu^-)$ pairing would be the most tightly bound;
assuming the $e^-e^+$ annihilate first, the remnant $\mu^+ \mu^-$
would then naturally form a true muonium atom, but at the larger
characteristic size of ordinary muonium.  For an even closer analogue
to the QCD system, one can imagine a hypothetical $\mu^-$ with charge
$-2|e|$.  It would still combine as $(\mu^+ e^-)(e^+ \mu^-)$, but each
component now has a net charge.

The case of QCD is of course more complicated.  However, one may note
that the $qq$ pair can couple to only two irreducible color
representations, ${\bm 3} \otimes {\bm 3} = \bar {\bf 3} \oplus {\bf
6}$ ({\it cf.} ${\bm 3} \otimes \bar {\bm 3} = {\bf 1} \oplus {\bf
8}$).  The binding of $qq$ or $\bar q q$ systems (call their component
color representations $R_{1,2}$) depends only on the quadratic Casimir
$C_2(R)$ of the product color representation $R$ to which the quarks
couple, according to $C_2 (R) - C_2 (R_1) - C_2 (R_2)$.  From this
simple rule, one finds the relative size of the couplings for a
quark-(anti)quark pair to be $\frac 1 3 (-8,-4,+2,+1)$ for $R = ({\bf
1},\bar {\bf 3},{\bf 6},{\bf 8})$, respectively; in particular, the
{\bf 1} is the unique attractive color channel for $\bar q q$, and the
$\bar {\bf 3}$ is the unique attractive color channel for $q q$.  It
is thus natural to model the tetraquark as a bound color-$\bar {\bf
3}$ diquark and a bound color-{\bf 3} antidiquark attracted by
color-Coulomb forces; this is the picture implicitly assumed in
Ref.~\cite{Maiani:2004vq}.

If one adopts the further ansatz that the dominant non-universal
interactions among the tetraquark components are spin-spin
interactions within each diquark, one obtains a very reasonable
explanation of the currently known $XY \! Z$
states~\cite{Maiani:2014aja}.  The primary limitation of this picture is
the fact that the (charged) isospin partners of the neutral $X,Y$
states have not yet been observed; for example, the $X(3872)$ in
Ref.~\cite{Maiani:2014aja} has the structure $\frac{1}{\sqrt{2}} ([c
q]_1 [\bar c \bar q]_0 + [c q]_0 [\bar c \bar q]_1)$, where $q$ is a
light quark and the subscript indicates the diquark spin.  No obvious
charged partners of the $X(3872)$ have been observed; however, in this
case, one can argue that $\bar q q$ is really the $I=0$ combination
$\frac{1}{\sqrt{2}} (\bar u u - \bar d d)$, and that the corresponding
$I=1$ states are somewhat higher in mass.  Furthermore, one can
explain~\cite{Karliner:2010sz} its tiny width and its ability to decay
into $J/\psi$ and either $\omega$ ($I=0$) or $\rho$ ($I=1$) by noting
that  the mass $m_{X(3872)}$ is slightly below $m_{J/\psi} + m_{\omega \, {\rm
peak}}$ but is very close to $m_{J/\psi} + m_{\rho \, {\rm peak}}$;   
the former channel is preferred by isospin conservation (QCD) but
suppressed by phase space, while the latter channel is favored by
phase space but suppressed by isospin-violating (QED) $\rho$-$\omega$
mixing.

As mentioned above, the conventional charmonium states are well fit
using solutions of a Schr\"{o}dinger equation with a nonrelativistic
potential $V(r)$.  In the most thorough recent
analysis~\cite{Barnes:2005pb}, the potential is taken to be
\begin{equation} \label{eq:Cornell}
V(r) = -\frac 4 3 \frac{\alpha_s}{r} + b r + \frac{32\pi\alpha_s}
{9m_c^2} \left( \frac{\sigma}{\sqrt{\pi}} \right)^3 \! \!
e^{-\sigma^2 r^2} {\bf S}_c \! \cdot {\bf S}_{\bar c} \, ,
\end{equation}
where $\alpha_s = 0.5461$, $b = 0.1425$~GeV$^2$, $m_c = 1.4797$~GeV,
and $\sigma = 1.0946$~GeV.  The origin of the color $-4/3$ factor as
the {\bf 3}-$\bar {\bf 3}$ coupling has already been noted.  Since the
first two terms in the potential $V(r)$ are determined solely by the
nature of the color field, one may use the same potential to describe
the interaction of the $\delta$-$\bar \delta$ interaction.  The third
(spin-dependent) term does depend upon the sources being fermionic,
but it is of short range, and its chief purpose is to accommodate
${}^3 \! S_1$-${}^1 \! S_0$ and ${}^3 \! P_J$ splittings.  Should the
$\delta$-$\bar \delta$ potential require a ${\bf S}_{qq} \! \cdot {\bf
S}_{\bar q \bar q}$ term, we note that it would still vanish for any
state consisting solely of terms for which at least one of $S_{qq}$ or
$S_{\bar q \bar q}$ vanishes; this condition is believed to
hold~\cite{Maiani:2014aja} for $X(3872)$, as well as for $Z(4430)^- =
\frac{1}{\sqrt{2}} ([c d]_1 [\bar c \bar u]_0 - [c d]_0
[\bar c \bar u]_1)$.

Let us carry the notion of the binding of compact diquarks to its
logical extreme, in which they can be considered as pointlike
color-triplet sources bound in the potential given by
Eq.~(\ref{eq:Cornell}).  To compute their eigenvalues, one must input
a value for the diquark mass $m_{cq}$.  We prefer to use a
determination that is independent of the nonrelativistic quark model,
so we use the value computed using QCD sum rules~\cite{Kleiv:2013dta}:
\begin{equation} \label{eq:mcq}
m_{cq} = 1.86 \pm 0.10 \ {\rm GeV,}
\end{equation}
for both spin-0 and spin-1 diquarks.  As long as $m_{cq} \! > \! m_c$,
one finds that the wave function becomes more compact: $\left< r
\right>_{J/\psi} = 0.39$~fm $>$ $\left< r \right>_{X_{1S}} = 0.31$~fm.
However, this use of the Schr\"{o}dinger equation now becomes suspect.
If one believes each diquark to have a size like that of a $D$ meson
(say, $\approx$ 0.5~fm), it is difficult to justify treating them as
pointlike compared to $\left< r \right>$.

\begin{figure}
\begin{center}
\includegraphics[width=3.4in]{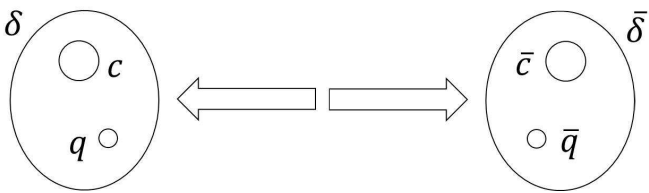}
  \caption{Diquark-antidiquark production.  The arrows indicate a
    color flux tube connecting the color-$\bar {\bf 3}$ diquark
    $\delta$ and the color-{\bf 3} antidiquark $\bar
    \delta$.}\label{Fig:diquark}
\end{center}
\end{figure}

\begin{figure}
\begin{center}
\includegraphics[width=3.35in]{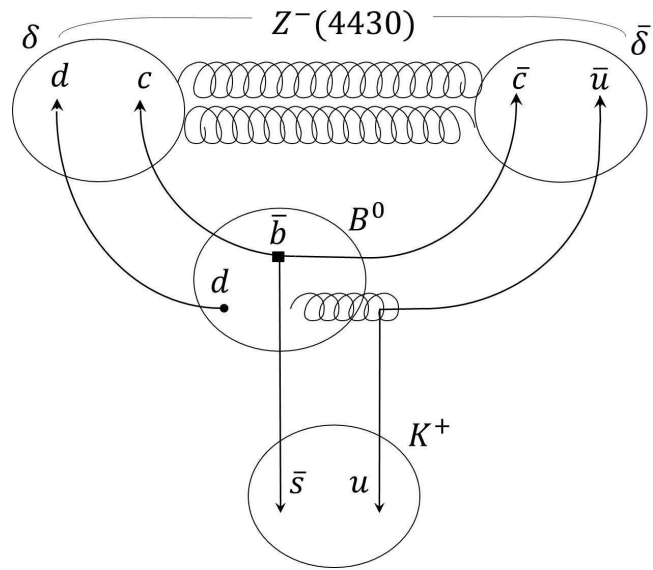}
  \caption{Illustration of the production of a spatially extended
  diquark-antidiquark state $\delta \bar \delta$ attracted by
  long-range color forces (indicated by gluon lines).  Here, the
  mechanism is illustrated for $B^0 \to Z(4430)^- K^+$, where the
  $\blacksquare$ indicates the $\bar b$-quark weak decay.}
\label{Fig:Z4430}
\end{center}
\end{figure}

We come at last to the central idea of this paper.  We propose that
the $XY \! Z$ states are not conventional nonrelativistic bound states
in the sense of solutions of a Schr\"{o}dinger equation for a static
potential, but are instead collective modes of a $\delta \bar \delta$
pair produced at a high relative momentum (Fig.~\ref{Fig:diquark}).
Were it not for confinement, the diquarks would fly apart as compact,
free (colored) mesons; instead, their large relative kinetic energy is
gradually converted into potential energy of the color flux tube
connecting them, an ``open-string hadron'' picture.  Eventually they
are brought relatively to rest after achieving a substantial
separation (Fig.~\ref{Fig:Z4430}).  We propose that this is the
dynamical physics underlying the formation of exotic tetraquark
states.  The tetraquark has an observably narrow width because of its
difficulty in hadronizing; if the $\delta \bar \delta$ pair were
replaced by a $\bar q q$ pair (again, which has the same $\bar {\bf
3}$-{\bf 3} color structure), the system would promptly fragment into
two or more mesons upon the creation of an additional $q^\prime \bar
q^\prime$ pair.

\begin{figure}
\begin{center}
\includegraphics[width=3.35in]{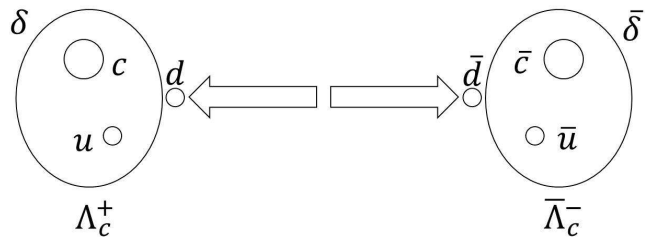}
  \caption{The baryonium system: Fragmentation of the color flux tube
    connecting the diquark-antidiquark pair $\delta \bar \delta$ into
    the lowest-lying baryon-antibaryon state, $\Lambda_c^+ \bar
    \Lambda_c^-$.}\label{Fig:baryonium}
\end{center}
\end{figure}

The corresponding scenario in the $\delta \bar \delta$ system would
produce a baryon-antibaryon pair (Fig.~\ref{Fig:baryonium}); in fact,
this system has long been considered in the literature and is called
{\it baryonium\/}~\cite{Rossi:1977cy}.  Should this $\delta$-$\bar
\delta$ picture truly hold for the $XY \!  Z$ states, one would expect
such a state with a dominant $\Lambda_c^+ \bar \Lambda_c^-$ decay mode
($\Lambda_c^+$ being the lightest charmed baryon) to appear very soon
after the threshold at 4573~MeV is passed; indeed, as noted in
Ref.~\cite{Cotugno:2009ys}, such a state has been observed, the
$X(4632)$.

\begin{figure}
\begin{center}
\includegraphics[width=3.4in]{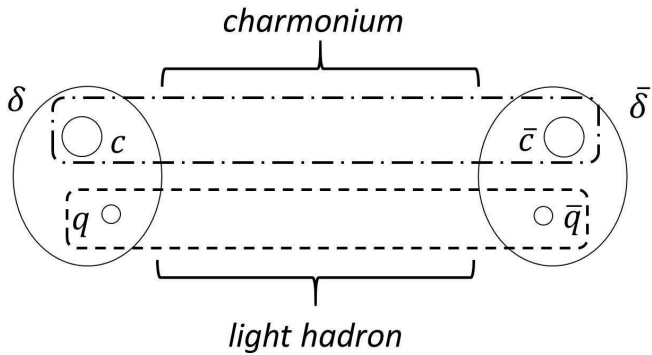}
  \caption{Illustration of the mechanism that suppresses the
  hadronization rate of diquark-antidiquark ($\delta \bar \delta$)
  extended states, due to the small wave-function overlap (indicated
  by dashed lines) at large spatial separation.}\label{Fig:tails}
\end{center}
\end{figure}

Ordinary fragmentation is not permitted below the $\Lambda_c^+ \bar
\Lambda_c^-$ threshold, and another hadronization mechanism must
occur.  One mechanism that has been identified in older literature
assumes that the quarks in $\delta$ meet their antiquark partners in
$\bar \delta$ (to form mesons) through a tunneling
process~\cite{Aerts:1979hn}; such a process would likely be very slow
and leads to near-stable tetraquark mesons.  However, we propose a
much simpler and quicker fate for the $\delta \bar \delta$ states:
Hadronization into charmonium and other mesons proceeds through the
large-$r$ tails of their wave functions stretching from the $\delta$
to the $\bar \delta$ (Fig.~\ref{Fig:tails}); the larger the
$\delta$-$\bar
\delta$ separation, the more suppressed the overlap integrals, hence a
more highly suppressed amplitude, and ultimately, a smaller width.

Because the extended $\delta \bar \delta$ system contains a great deal
of color energy, it hadronizes (albeit with a small width) almost as
soon as a threshold for creating hadrons with the same quantum numbers
is passed, and final states with the smallest number of particles
should dominate.  In the case of the $J^{PC} =1^{++}$ $X(3872)$, the
first such isospin-conserving threshold is $J/\psi + \omega$.  The
combination of the small $\left< J/\psi \; \omega \left|
\right. \! \delta \bar \delta \right>$ overlaps, small phase space,
and the role of isospin violation discussed above can conspire to give
the $X(3872)$ its surprisingly small width.  In this picture, one
expects to find a $\delta \bar \delta$ resonance just beyond the
threshold for the lowest charmonium-light meson threshold that has the
same quantum numbers; for example, an isotriplet with $J^P = 1^+$ is
expected not far above $m = m_{J/\psi} + m_{\rho^+} = 3872$~MeV or the
$I=1$ channel of $D \bar D^*$ ($\simeq 3876$~MeV), and indeed, BESIII
finds~\cite{Yuan:2014rta} a $1^+$ state of mass $3883.9 \pm 1.5 \pm
4.2$~MeV [$Z_c(3885)$, which is believed to be the $Z(3900)$ coupled
to the $D \bar D^*$-meson modes].  The means by which a resonance can
synchronize with a threshold (the so-called {\it cusp effect}) is
discussed in Ref.~\cite{Bugg:2008wu}, and interesting effects that
occur when related channels couple near threshold is discussed in
Ref.~\cite{Voloshin:2006pz}.

However, not every $XY \! Z$ state in this picture needs to lie near a
threshold.  In the diquark model of Ref.~\cite{Maiani:2014aja},
multiplets arise from different orientations of the spins within
$\delta$ and $\bar \delta$, as well as their combination to form the
$J^P$ of the full $\delta \bar \delta$ state.  Once one state is
formed from the appearance of a threshold, so do several others.  In
addition, the color flux tube that connects the $\delta$ and $\bar
\delta$ also can be excited; in Ref.~\cite{Maiani:2014aja}, the
$J^{PC} = 1^{--}$ $Y$ states are identified as its orbital $L=1$
excitations.  Furthermore, the flux tube can have longitudinal
excitations, which were identified in Ref.~\cite{Maiani:2014aja} as
radial modes; in particular, the $Z(4430)$ is identified as the $2S$
partner to the $1S$ $Z(3900)$.  This identification is made in part
due to the strong preferential couplings of $Z(3900) \to J/\psi$ and
$Z(4430) \to \psi(2S)$.

The preferential coupling of some of the $XY \! Z$ states to the
$\psi(2S)$ has been a longstanding mystery.  After all, both $J/\psi$
and $\psi(2S)$ are $J^P = 1^{--}$ states, and one naively expects any
state that can decay to $\psi(2S)$ can also couple to the much lighter
$J/\psi$.  Let us now perform a simple calculation to show how our
picture leads to a natural observed preference of $B^0 \to Z(4430)^-
K^+ \to \psi(2S) \pi^- K^+$.  Using as inputs $m_{B^0}$, $m_{K^+}$,
and $m_{Z(4430)}$,
transforming into the c.m.\ frame of the $\delta$-$\bar \delta$ pair,
and using $m_{cq}$ from Eq.~(\ref{eq:mcq}), one obtains the initial
$\delta$-$\bar \delta$ pair kinetic energy $T_{cq}$.  Using
Eq.~(\ref{eq:Cornell}) and its numerical inputs, all of the $T_{cq}$
kinetic energy converts into potential energy at the large
$\delta$-$\bar \delta$ separation $r_Z = 1.16$~fm.  In comparison, the
corresponding value for the $X(3872)$ is $r_X = 0.56$~fm.

The natural figures of merit we use to compare the relative likelihood
of $Z(4430) \to J/\psi + \pi$ and $Z(4430) \to \psi(2S) + \pi$
transitions are the ratios of probability densities of the $1S$ and
$2S$ states evaluated at $r_X$ and $r_Z$.
Using Eq.~(\ref{eq:Cornell}), one finds the values
$|\Psi_{1 \! S}(r)/\Psi_{2S}(r)|^2_{r=r_X} = 2.41 : 1$ and $|\Psi_{1
\! S}(r)/\Psi_{2S}(r)|^2_{r=r_Z} = 1 \! : \! 75.6$.  The preference of
the decay $Z(4430) \to \psi(2S) + \pi$ [and $Z(3900) \to J/\psi +
\pi$] is thus very natural in this picture.  Clearly, one may consider
any suitable weighting functions about $r_{X,Z}$ and obtain comparable
results.  The anticipated preferential decay of the tetralepton $\mu^+
e^- \mu^- e^+$ to positronium plus a large-size $\mu^+ \mu^-$ atom has
a similar origin.

Why are tetraquark states obvious in the charm (and likely
bottom~\cite{Karliner:2013dqa}) sector but not in the light-quark
sector?  Again, we believe that our picture provides guidance, based
on the energy scales of the tetraquark states.  While the scalars
$a_0(980)$ and $f_0(980)$ may also be $\delta \bar \delta$
states~\cite{Maiani:2004uc} (and indeed have been argued to be
tetraquarks for a long time~\cite{Jaffe:1976ig}; note also recent
work~\cite{Stone:2013eaa,Aaij:2014siy} to measure their tetraquark
content), it is not clear that their $\delta$ and $\bar \delta$
components are sufficiently compact that they do not overlap and
instantly mix with a meson-meson configuration.  In the case of decays
with heavier quarks like charm, we have seen that the large energy
release makes the $\delta$-$\bar \delta$ separation sufficiently large
to distinguish them.

This calculation is decidedly quite crude.  We have used a classical
turning point for the potential, which can be improved via the
Wentzel-Kramers-Brillouin approximation.  We have used one particular
determination of the diquark mass (which in turn is treated as a
compact quasiparticle), which of course can be varied.  We have used
the phenomenological nonrelativistic potential $V(r)$ of
Eq.~(\ref{eq:Cornell}) to represent confinement physics and ignored
that relativistic systems should be described in terms of helicities
rather than spins; however, one may use instead a lattice-based
determination, or the result of an AdS/QCD calculation which uses
fully covariant front-form dynamics, and obtain comparable
results~\cite{Trawinski:2014msa}.

In conclusion, we have presented a dynamical picture to explain the
nature of the exotic $XY \! Z$ states based on a diquark-antidiquark
open-string configuration.  Our model provides a natural explanation
of why some, but not all, of the states lie very close to hadronic
thresholds, and why some of the states prefer to decay to excited
charmonium states.  Although this picture is only semi-quantitative,
we believe it provides a good starting point for understanding the
structure, formation, and decay of these exotica.

{\it Acknowledgments.}  This work was supported by the U.S.\
Department of Energy under Grant No.\ DE-AC02-76SF00515 (S.J.B.); by
the Korea Foundation for International Cooperation of Science \&
Technology (KICOS) and the Basic Science Research Programme through
the National Research Foundation of Korea (2010-0011034) (D.S.H.); and
by the National Science Foundation under Grant No.\ PHY-1068286
(R.F.L.).  In addition, D.S.H.\ thanks H.~Son for technical
assistance, and R.F.L.\ thanks T.~Cohen, L.~Haibo, and T.~Skwarnicki
for enlightening exchanges.


\end{document}